\documentclass[twocolumn,
 superscriptaddress,amsmath,amssymb,nofootinbib]{revtex4}
\usepackage{amsfonts}
\usepackage{amsmath}
\usepackage{bbm}
\usepackage{amssymb}
\usepackage{graphicx}%
\usepackage{footmisc}
\usepackage{bm}
\usepackage{braket}
\usepackage{CJK}
\usepackage{hyperref}
\usepackage{bbm}
\usepackage{cleveref}
\usepackage{longtable,booktabs}

\crefname{equation}{Eq.}{Eqs.}

\newcommand{\abs}[1]{\lvert#1\rvert}
\newcommand{\norm}[1]{\lVert#1\rVert}
\newcommand{\Tr}{\text{Tr}}

\begin{document}
\begin{CJK*}{UTF8}{gbsn}
\title{Quantifying non-Markovianity via  conditional mutual information }

\author{Zhiqiang Huang (黄志强)}
\email{hzq@wipm.ac.cn}
\affiliation{Innovation Academy for Precision Measurement Science and Technology\footnote{State Key Laboratory of Magnetic Resonance and Atomic and Molecular Physics, National Centre for Magnetic Resonance in Wuhan, and Wuhan Institute of Physics and Mathematics.}, CAS, Wuhan 430071, China}
\author{Xiao-Kan Guo~(郭肖侃)}
\affiliation{South China Normal University, Shanwei, China}

\date{\today}

\begin{abstract}
    In this paper, we study measures of quantum  non-Markovianity  based on the conditional mutual information. We obtain such measures by considering multiple parts of the total environment such that the conditional mutual information can be defined in this multipartite setup.
    The benefit of this approach is that the conditional mutual information  is closely related to recovery maps and Markov chains; we also point out its relations with  the change of distinguishability. We study along the way the properties of  leaked information which is the conditional mutual information that can be back flowed, and we use this leaked information to show that the correlated environment is necessary for nonlocal memory effect. 
\end{abstract}


\maketitle

\section{Introduction}
Open quantum systems are ubiquitous in the  realistic quantum world. The Markovian approximation allows us to obtain an exact dynamical description of the  open quantum dynamics  via the
 Lindblad-Gorini-Kossakowski-Sudarshan  master equation. Beyond this approximation, we have the non-Markovian quantum dynamics with memory effects whose mathematical descriptions  remain elusive. Although there have been a wide variety of approaches to the non-Markovian dynamics, no consensus is reached so far. See, e.g. \cite{BHPV16,VA17} for recent reviews.

 To characterize the differences between the non-Markovian open quantum processes and the Markovian ones, one can  define the non-Markovianity measures as the mathematical characterizations other than the master equations. The  attempts to quantify the quantum non-Markovianity, including directly defining the characteristic measures \cite{Bre12,RHP14} and by applying the  quantum resource theory \cite{CG19}. 
  Currently, two simple  typical definitions of quantum Markovian processes are the completely positive (CP) divisibility of dynamical maps  \cite{RHP10}, and the non-existence of the information backflow  under dynamical maps \cite{BLP09}; the corresponding non-Markovianity measures are known respectively as the Rivas-Huelga-Plenio (RHP) measure and the Breuer-Laine-Piilo (BLP) measure.  A comparison of these two typical measures can be found, for example, in  \cite{DKR11}.
Notice that the no-backflow condition is more general than CP-divisibility, because it is definable even if there are classical memories \cite{BMHH20}.

In defining the non-Markovianity measures, it is  desirable to take into account all possible  memory effects.
 In fact, the non-Markovianity measure based on general (both quantum and classical) correlations can be defined via the  quantum mutual information \cite{LFS12}, which we call the  Luo-Fu-Song (LFS) measure. The recent work \cite{DJ20} shows that it is possible to find a one-to-one correspondence between the CP-divisibility and the condition of no correlation backflow. Therefore, the non-Markovianity measures based on correlations, such as the LFS measure, can evade the distinction made in \cite{BMHH20} and present a clear characterization of non-Markovianity.

All these measures of quantum non-Markovianity are defined for  open quantum systems (and their   dynamics); the structures of  environment are  hidden in the reduced descriptions of the open quantum systems, which hinders further  identifications of the origins of memory effects. It is an interesting question   that  how the structures of environment affect non-Markovianity, especially when the initial system-environment state is correlated.

In this paper, we study the effects of the structured environment on the non-Markovianity of the open quantum system. 
We first find an equivalent form  of the LFS measure in terms of the quantum conditional mutual information defined in the system+ancillary+environment setup. Using this new form of non-Markovianity measure, we  study how parts of the environment affect the memory effects by considering the conditional mutual information with respect to the sub-environments obtained by the chain rule.
In addition, we can keep track of the system-(part-of)-environment correlations. In doing so, we try to find the possible origin of memory effect from the perspective of parts of the environment, which is not easy to study if one only focuses on the open system.

In \cref{s22},  we  show the general  relation between LFS measure  and the change in the distinguishability of states in a way similar to the BLP measure.  We then present in \cref{s2} a  reformulation of  the LFS  non-Markovianity measure based on quantum conditional mutual information, i.e. $\mathcal{N}_1$ (Eq. \eqref{NMM}). Using this new form $\mathcal{N}_1$, we discuss the relations between the (Petz) recoverability and the distinguishability  used in defining the BLP measures;  we exploit  the leaked information, the quantum mutual information that can backflow into the system, which  explicitly contains the impact of the parts of  environment.  
The leaked information can be applied to characterize the nonlocal memory effect, and we show numerically that the classically correlation in the environment may not give rise to the nonlocal memory effect.
\cref{CO} concludes with some outlooks.

\section{Non-Markovianity measure from mutual information}\label{s22}
Consider an open quantum system $S$ interacting with an environment $E$; $S$ and $E$ form a closed total system with unitary evolution. The dynamical evolution of the  state $\rho_S$ of the system $S$ is  represented by a completely positive trace preserving (CPTP) map $\Lambda_t$ such that $\rho_S(t)=\Lambda_t\rho_S(0) $. The Markovian dynamical maps in the RHP sense are the CP-divisible maps, i.e. $\Lambda_t=\Lambda_{t,r}\Lambda_r$ for $r\leqslant t$.

In order to characterize the correlations in $\rho_S$, we make $\rho_S$ into a bipartite system $\rho_{SA}$ by adding   an ancillary system $A$ that evolves trivially by the identity map $\mathbbm{1}$.  Then $\rho_{SA}$ evolves as
$\rho_{SA}(t)=(\Lambda_t\otimes\mathbbm{1})\rho_{SA}(0)$.
The total correlations shared by $S$ and $A$ is quantified by the quantum mutual information $I(\rho_{SA})=I(S:A)=S(\rho_S)+S(\rho_A)-S(\rho_{SA})$ where $S(\rho)$ is the von Neumann entropy.
Since $I(\rho_{SA})$ is   monotonically decreasing, i.e. $dI(\rho_{SA}(t))/dt\leqslant0$, under the Markovian local operation $\Lambda_t\otimes\mathbbm{1}$, the increasing part of the mutual information can be exploited to define the LFS non-Markovianity measure \cite{LFS12} for a dynamical map $\Lambda$,
\begin{equation}\label{NMML}
    \mathcal{N}_{\text{LFS}}(\Lambda)=\sup_{\rho_{SA}}\int_{\frac{d}{dt}I(\rho_{SA}(t))>0}\frac{d}{dt}I(\rho_{SA}(t))dt
\end{equation}
where the sup is over all those $\rho_{SA}$. Notice that the derivative $\frac{d}{dt}I(\rho_{SA}(t))$ in effect witnesses the quantum non-Markovianity (cf. the recent paper \cite{DJBBA20}); but the LFS measure $\mathcal{N}_{\text{LFS}}$ is a quantification of quantum non-Markovianity, i.e. the integration gives back an informational quantity, rather than just witnessing it.

 Recall also the BLP non-Markovianity measure \cite{BLP09},
\begin{equation}\label{BLP}
    \mathcal{N}_{\text{BLP}}(\Lambda)=\sup_{\rho,\tau}\int_{\frac{d}{dt}D(\Lambda_t\rho,\Lambda_t\tau)>0}\frac{d}{dt}D(\Lambda_t\rho,\Lambda_t\tau)dt
\end{equation}
where $D(\rho,\tau)$ measures the distinguishability of two states. More generally, one  retains the interpretation of \eqref{BLP} as the distinguishability of  states under quantum dynamical maps, even if  the trace distance by other distance measures, for example, the fidelity \cite{VMPBP11}.
In the following, we will consider another measure of distinguishability related to the quantum conditional mutual information.

 In many examples considered in \cite{LFS12}, the LFS measure is consistent with the BLP measure. 
 It is known, however, from e.g. \cite{,ALDPP14,CLCC15}, that there is a hierarchy of non-Markovianity measures: LFS$\leqslant$BLP$\leqslant$RHP, meaning that the LFS measure detects less non-Markovianity than the BLP measure.
 Here, through a general proof,we show that the change $\delta  I(A:S)$ is indeed related to the distinguishability of states under dynamical maps (or distinguishability, for short).

The quantum mutual information $I(A:S)$ can be expressed in terms of the quantum relative entropy as \cite{Ved02}
\begin{equation}\label{6}
I(S:A)=S(\rho_{SA}||\rho_S\otimes\rho_A).
\end{equation}
By this argument, we know that the change $\delta I(A:S)$ is the same as the change  $\delta S(\rho_{SA}||\rho_S\otimes\rho_A)$. To relate  $\delta S(\rho_{SA}||\rho_S\otimes\rho_A)$ to the distinguishability of states, let us consider the optimal pair of states $\rho^1_S$ and $\rho^2_S$, i.e. the pair of states for which   the maximum in $\mathcal{N}_{\text{BLP}}(\Lambda)$ is attained.
According to \cite{WKLPB12}, $\rho^1_S$ and $\rho^2_S$ are orthogonal  states on the boundary of the state space. Then we can construct  a
correlated initial system-ancillary state as the superposition of orthogonal states
\begin{equation}\label{TSIO}
    \rho_{SA}^{op}=\frac{1}{2} (\rho^1_S\otimes \Pi^1_A+\rho^2_S\otimes \Pi^2_A),
\end{equation}
where the $\Pi_A^{1,2}$ are the projection operators satisfying $\Pi_A^1\Pi_A^2=0$.
Under time evolution, the projectors $\Pi_A^{1,2}$ can be taken as time-independent, i.e.
\begin{equation}\label{TSIOT}
    \rho_{SA}^{op}(t)=\frac{1}{2} (\rho^1_S(t)\otimes \Pi^1_A+\rho^2_S(t)\otimes \Pi^2_A).
\end{equation}
The corresponding uncorrelated product state is
\begin{equation}
    \rho_{S}(t)\otimes\rho_A=\frac{1}{2} (\rho^1_S(t)+\rho^2_S(t))\otimes\frac{1}{2}(\Pi^1_A+ \Pi^2_A).
\end{equation}
Since $\Pi_A^1\Pi_A^2=0$, we have
\begin{equation}\label{99}
       S(\rho_{SA}^{op}(t)||\rho_S(t)\otimes\rho_A)=\log 2 \times D_{\text{tele}}(\rho^1_S(t),\rho^2_S(t)),
\end{equation}
where 
\begin{equation}
    D_{\text{tele}}(\rho,\sigma)=(S_{1/2}(\rho||\sigma)+S_{1/2}(\sigma||\rho))/2
\end{equation}
is the quantum Jensen-Shannon divergence and $S_a(\rho||\sigma)$ is the  $a$-telescopic relative entropy \cite{Aud11}
\begin{equation}\label{teleRE}
    S_a(\rho||\sigma)=\frac{1}{-\log a}S\bigl(\rho||a\rho+(1-a)\sigma\bigr),~a\in(0,1).
\end{equation}
See \cref{AnnB} for the derivation of \eqref{99}. 

The telescopic relative entropy can be used as the distance measure between quantum states, so that we have a special type of the BLP non-Markovianity measure
\begin{equation}\label{teleBLP}
    \mathcal{N}_{\text{tBLP}}(\Lambda)=\sup_{\rho,\tau}\int_{\frac{d}{dt}D_{\text{tele}}(\Lambda_t\rho,\Lambda_t\tau)>0}\frac{d}{dt}D_{\text{tele}}(\Lambda_t\rho,\Lambda_t\tau)dt,
\end{equation}
where the sup is obtained for the optimal state pair \eqref{TSIO}.

Now we  have at least
\begin{equation}
    \mathcal{N}_{\text{tBLP}}(\Lambda)=\frac{1}{\log 2} \int_{\frac{d}{dt}I(\rho_{SA}(t))>0}\frac{d}{dt}I(\rho_{SA}(t))dt |_{\rho_{SA}=\rho_{SA}^{op}},
\end{equation}
but the optimal state pair \eqref{TSIO} might not give rise to the supremum of $\mathcal{N}_{\text{LFS}}$. Since   $\mathcal{N}_{\text{LFS}}=0$ always implies $\mathcal{N}_{\text{tBLP}}=0$ by definition, we see that
  $\mathcal{N}_{\text{tBLP}}$ doesn't detect more Markovianity than $\mathcal{N}_{\text{LFS}}$, or equivalently LFS$\leqslant$tBLP. This is of course consistent with the hierarchy of non-Markovianity measures.

As a consequence, the measure $\mathcal{N}_{\text{LFS}}(\Lambda)$ quantifies in effect the distinguishability of quantum states under dynamical maps, if we choose the $\rho_{SA}$ in the special form of  \eqref{TSIO}. 
 A non-Markovian quantum process implies
the increasing of distinguishability, i.e.  $\delta  S(\rho_{SA}||\rho_S\otimes\rho_A)>0$, whereas for Markovian quantum process with (CP-divisible) CPTP map $\Lambda$, one has
\[
     \delta S_a(\rho||\sigma)=  S_a( \Lambda(\rho)|| \Lambda(\sigma))-S_a(\rho||\sigma) \leqslant 0,
\]
whereby one obtains
$    \delta S(\rho_{SA}||\rho_S\otimes\rho_A)  \leqslant 0$. This behavior is consistent with the properties of  other types of relative entropies that have been used to quantify distinguishability, e.g. \cite{LL19,npj}.

\section{An equivalent measure via conditional mutual information}\label{s2}
Let us turn to the quantum conditional mutual information in the ``system+ancillary+environment'' setup. Since the open system dynamics is given by the  unitary evolutions of the closed system-environment states and unaffected by the trivial evolution of $A$, the  quantum mutual information $I(A:SE)$ between the ancillary state and the system-environment total state should be time-independent (otherwise the exchange in correlations will make the system-environment total system  open). It is easy to show that
\begin{equation}\label{2}
    I(A:SE)=I(A:E|S)+I(S:A),
\end{equation}
where the quantum conditional mutual information is
\begin{equation}
    I(A:E|S)=S(\rho_{AS})+S(\rho_{SE})-S(\rho_{S})-S(\rho_{ASE}).
\end{equation}
Due to the strong subadditivity of  von Neumann entropy, we have $I(A:E|S)\geqslant0$.

\begin{figure}[t]
    \centering
    \includegraphics[width=0.5\textwidth]{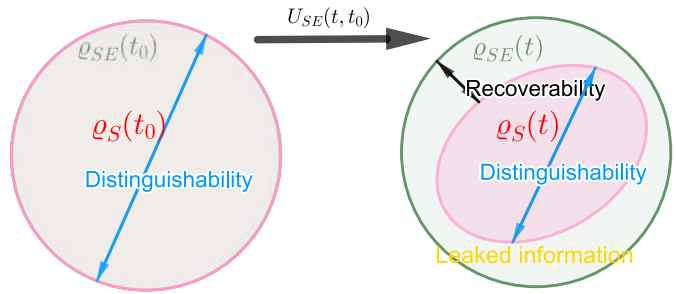}
    \caption{The global unitary evolution of the system+environment will not change the state space and the global distinguishability, but the state space and the distinguishability and recoverability of the open system will decrease under the dynamical evolution. }\label{f1}
    \end{figure}

Now consider the time-derivative of Eq. \eqref{2}. Since $I(A:SE)$ is time-independent, i.e.  $dI(A:SE)/dt=0$, we have that $dI(S:A)/dt$ and $dI(A:E|S)/dt$ have the same magnitude but opposite signs. From \eqref{NMML} we know that for non-Markovian quantum processes, $dI(S:A)/dt>0$, which entails $dI(A:E|S)/dt<0$. Then, in analogy to the LFS measure \eqref{NMML}, we  define the following non-Markovianity measure for a dynamical map $\Lambda$
\begin{equation}\label{NMM}
    \mathcal{N}_1(\Lambda):=  \sup_{\rho_{SA}}\int_{\frac{d}{dt} I(A:E|S)<0} \abs{ \frac{d}{dt} I(A:E|S)}dt
\end{equation}
where the sup is still over the system-ancillary states $\rho_{SA}$. Now, we briefly discuss the physical meaning of the quantum conditional mutual information.  In the resource theory of non-Markovianity based on the Markov chain condition \cite{CG19}, the non-vanishing magnitude of the quantum conditional mutual information indicates the violation of Markovianity. Therefore, the measure  $\mathcal{N}_1(\Lambda)$ relates in an interesting way two Markovian conditions.  Besides that, the quantum conditional mutual information is related to recoverability (cf. \cref{AnnA}) and distinguishability (cf. \cref{s22}), the decrease of which indicate the   ``contracts'' in the state space of system indeed.  Hence, it also implies the change in the volume of  quantum state space, which is another characterization of quantum non-Markovianity  \cite{LPP13}. Fig.\ref{f1} gives a brief illustration of these relationships.

 This new form of measure \eqref{NMM} is dependent on $E$, but it  actually depends only on those parts  that interact with the system. To see this, suppose the environment $E$ consists of two sub-environments $E_1$ and $E_2$, and $E_1$ interacts with $S$ while $E_2$ does not interact with $S$. Then under the local unitary $U_{SE_1}$, $I(A:E_2|SE_1)$ is unchanged, so that by the chain rule \eqref{B6},
\begin{equation}\label{SubE}
    \delta I(A:E_1E_2|S)= \delta I(A:E_1|S).
\end{equation}
Consequently, the $E_2$ will not contribute to the measure \eqref{NMM}.\footnote{The similar  can be said for subsystems: only those subsystems interacting with the environment are of interest.}
In this particular case, we have equivalently
\begin{equation}\label{N1part}
    \mathcal{N}_\text{part}(\Lambda)=  \sup_{\rho_{SA}}\int_{\frac{d}{dt} I(A:E_1|S)<0} \abs{ \frac{d}{dt} I(A:E_1|S)}dt.
\end{equation}
On face of it,  in \eqref{NMM},  the previously considered system-environment correlation $I(S:E)$ is now changed to the ancillary-environment correlation conditioned on the system, $I(A:E|S)$. If we understand the ancillary as the tool for representing the quantum coherence of the system by the quantum correlation between the system and the ancillary, changing to $I(A:E|S)$ still means that the direction of information backflow in into the system. (Since we have assumed $I(A:SE)=0$, the changes in the correlation between $A$ and $E$ should be balanced by the changes in the correlation between $A$ and $S$.) 

It is also straightforward to generalize the $\mathcal{N}_1$ to an RHP-type measure. We perform such a generalization in \cref{BBB}.

Although this $ \mathcal{N}_1(\Lambda)$ is obviously equivalent to the LFS measure $ \mathcal{N}(\Lambda)$, the consideration of parts of environment allows us to exploit the properties of 
the quantum conditional mutual information (cf. \cref{AnnA}), as we now discuss in the following subsections.  


\subsection{Leaked information and multiple environments}\label{s2a}
As is shown in \cref{AnnA}, the exact recovery of a quantum channel is given by the vanishing of the conditional mutual information $I(A:E|S)=0$, the Markov chain condition. When the evolution of the open system is non-Markovian under the definition of LFS measure, we should have both
  \begin{equation}\label{41}I(A:E|S)\neq0\quad\text{and}\quad\frac{d}{dt} I(A:E|S)<0.\end{equation}
Since $I(A:E|S)$ can  be used as a measure and the decreasing of  $I(A:E|S)$ is the sufficient condition for non-Markovianity, here we call it  {\it leaked information}. 
Notice that the leaked information thus defined has already appeared in \cite{loss} under the name of {\it quantum loss}; however, \cite{loss} studies the case in which the total system-ancillary-environment is a pure state and ignores the dependence of $I(A:E|S)$ on the environment. Here, we focus on the effects of structured environment.
Suppose the map $\Lambda$ is induced by single unitary evolution $U_{SE_i}$, then according to \cref{SubE}, leaked information $I(A:E_i|S)$ gives an upper  bound for  $ \mathcal{N}_1(\Lambda)$, i.e., the leaked information limits the backflow of information. In order to further understand leaked information and the environmental effects, let us first turn to two properties of quantum conditional mutual information in the multiple-environment scenario.

Firstly, we show that there are entanglement phenomena in leaked information.
According to the discussions around \cref{SubE} and in \cref{LEII}, we see that the leaked information can be quantified by $I(A:E_d|S)$ where $E_d$ is the sub-environment that directly interacts with the system. Suppose the environment $E$ consists of two parts $E_1$ and $E_2$, each of which can interact with the system. Then by the chain rule we have
\begin{align}
    I(A:E_1E_2|S)=  I(A:E_1|S)+  I(A:E_2|S)\notag \\
    - I(E_1;E_2;A|S),\label{14}
\end{align}
where
\begin{equation}\label{II}
  I(E_1;E_2;A|S)= I(E_1:E_2|S)-I(E_1:E_2|SA).
\end{equation}
The second line of \eqref{14} is similar to the quantum interference term showing the interplay between the two sub-environments $E_1$ and $E_2$. In fact, the quantum conditional mutual information contains the quantum entanglement as well as other types of correlations.

By the definition of  squashed entanglement \cite{CW04}, 
\begin{equation}
E_{sq}(A:B)=\frac{1}{2}\inf_{\omega_{ABE}}\bigl\{I(A:B|E)|\rho_{AB}=\text{Tr}_E(\omega_{ABE})\bigr\}
\end{equation}
we know that the squashed entanglement  $E_{sq}(A:E)$ is half of the infimum of $I(A:E|X)$, so the leaked information contains the (squashed) entanglement between $A$ and $E$. The squashed entanglement is monogamous
\begin{equation}
    E_{sq}(A:E_1E_2)\geq  E_{sq}(A:E_1)+ E_{sq}(A:E_2),
\end{equation}
which means that  $ I(E_1;E_2;A|S)$ can be negative. In such cases, there exists non-local leaked information.

Secondly, it is possible for  $I(A:E|S)$ to contain classical correlations. We notice the following properties of quantum conditional mutual information: (i) positivity; (ii) invariance under the addition of sub-environments in the tensor-product form, $I_{A:EE'|S}(\rho_{ASE}\otimes\rho_{E'})=I_{A:E|S}(\rho_{ASE})$; (iii) invariance under the local unitary transformations on $S+E$,
\[
    I_{A:E|S}(U_S\otimes U_E\rho_{SEA}U_S^\dagger\otimes U_E^\dagger)=  I_{A:E|S}(\rho_{SEA}).
\]
If the leaked information can be broadcast among multiple sub-environments
\begin{equation}
    \rho_{ASE_1}\to U_{E_1E_2\dots E_N} (\rho_{ASE_1}\otimes  \rho_{E_2\dots E_N} )U^\dagger_{E_1E_2\dots E_N}\equiv\sigma,
\end{equation}the multiple sub-environments would have the same amount of leaked information, i.e. $I_{A:E_i|S}(\sigma)=I_{A:E_1|S}(\rho)$. Since the broadcast can be achieved with the addition of sub-environments and local unitary transformations,  we have 
\begin{equation}\label{19}
    I_{A:E_1E_2\dots E_N|S}(\sigma)= I_{A:E_1|S}(\rho)= I_{A:E_i|S}(\sigma),
\end{equation}
where we have used (ii,iii) properties of quantum conditional mutual information.
By comparing \eqref{14} and \eqref{19}, we see that for \eqref{19} to hold, the $I(E_1;E_2;A|S)$ is positive. This means  $I(A:E_2|S)$ is  redundant (or repeated) leaked information, which can be eliminated by $I(E_1;E_2;A|S)$. From the perspective of resource theory, the redundant leaked information is non-resourceful.

Since non-local leaked information gives negative $ I(E_1;E_2;A|S)$,  the leaked information of sub-environment will be  suppressed. And the term $ I(E_1;E_2;A|S)$ will also eliminate the non-resourceful leaked information. Hence, these two properties of leaked information place significant restrictions on the backflow of information. 
They may help us find out why the backflow of  information in large environment is difficult.
\subsection{Nonlocal memory effect and correlated environment}
In this subsection, we use the leaked information to study the effect of environment correlation on the nonlocal memory effect of the open quantum system and also on the measure $\mathcal{N}_1$.

Recall that in \cite{nlm} the initial total  state $\rho_{S_1S_2}\otimes\rho_{E_1E_2}$ is evolved by two unitaries $U_{S_1E_1}$ and $U_{S_2E_2}$ during different periods of time, and it is found that when there exists quantum entanglement between $E_1$ and $E_2$ the open system $S_1S_2$ follows a non-Markovian evolution, even if the local dynamics of $S_1$ and $S_2$ is Markovian.
While if $E_1$ and $E_2$ are uncorrelated then $S_1S_2$ has Markovian open dynamics. 

Notice that, although we have \cref{SubE} under the  unitary $U_{S_1E_1}$, generally speaking 
$\delta I(A:E_1|S_1S_2)\neq \delta I(A:E_1|S_1)$. But it is easy to show that  $\delta I(A:E_1E_2|S_1S_2)=\delta I(A:E_2|S_1S_2)$ under the  unitary $U_{S_2E_2}$.
Now we would like to have that, if initially the sub-environments $E_1$ and $E_2$ are uncorrelated, then there is no leaked information in $E_2$ about $S_1S_2$ after the first unitary $U_{S_1E_1}$, namely 
$I(A:E_2|S_1S_2)=0$. Indeed, the initial total  state $\rho_{AS_1S_2}\otimes\rho_{E_1}\otimes\rho_{E_2}$ means that there is no  leaked information
\begin{equation}
    \sup_{\rho_{SA}} I(A:E_2|S_1S_2)=0
\end{equation}
at the initial moment, and it is expected that it remained zero under $U_{S_1E_1}$ if there is no leaked information in $E_2$.  
We can therefore use the leaked information $I(A:E_i|S)$ to characterize the appearance of nonlocal memory effect: If $I(A:E_2|S)>0$ after the local unitary $U_{S_1E_1}$,  then it means that $U_{S_1E_1}$ generates the leaked information in $E_2$, which is a necessary condition for the existence of nonlocal memory effect; after the second unitary $U_{S_2E_2}$,  we know from \cref{SubE} that $\delta I(A:E_2|S)$ equals the total change in the leaked information, whereby the decease in $ I(A:E_2|S)$ implies the {\it nonlocal memory effect}. 

Next, we consider an example from \cite{nlm,WB13} but  study the case in which the sub-environments are classically correlated.  We will numerically show that the classically correlated environment cannot give rise to 
the nonlocal memory effect (as characterized by the leaked information).

The model consists of two qubits as the system and two multimode bosonic baths as the environment; the total Hamiltonian reads
\begin{equation}\label{ttH}
    H=\sum_{i=1}^2\Bigl(\epsilon_i\hat{\sigma}_z^i+\sum_k\omega_k^i \hat{b}_k^{i\dagger}\hat{b}_k^{i}+\chi_i(t)\sum_k (g_k^i\hat{\sigma}_z^i  \hat{b}_k^{i\dagger}+H. c.)\Bigr),
\end{equation}
where the first term is the system Hamiltonian of two qubits, the $\hat{b}_k^{i\dagger}$ is the creation operator of the $k$-th mode of the $i$-th bath so that the second term is the environment Hamiltonian, and the third term means the interaction with couplings $g_k^i$ and $\chi_i(t)=\Theta(t-t_i^s)\Theta(t_i^f-t)$.  In the interaction picture, the evolution generated by $H$ is  given by
  \begin{equation}\label{UE}
      U(t)=e^{-i H_0 t}d(t) \exp\Bigl\{\sum_{i,k}(\beta_k^i(t)\hat{\sigma}_z^i \hat{b}_k^{i\dagger} -H. c.)\Bigr\},
  \end{equation}
 where $d(t)$ is a phase factor and
 $\beta_k^i(t)=\frac{g_k^i}{\omega_k^i}e^{i \omega_k^i t_i^s}(1-e^{i\omega_k^i\int _0^t ds \chi_i(s))})$. Moreover, we assume that different modes of the environment are independent of each other, so that the environment state is a product state,
\begin{equation}\label{modin}
    \rho=\prod_k \rho_k=\exp(\int \log \rho_k  dk).
\end{equation}

Instead of entangled environment state, we consider here the environment state with only classical correlation
\begin{equation}\label{242424}
    \rho_u^k= (1-u^2)\sum_{n=0}^\infty u^{2n} \Pi_n^{1,k}\otimes  \Pi_n^{2,k},
\end{equation}
where $\Pi_n^i=\ket{n}_i\bra{n}$ and $u=\tanh r$ is a parameter (which is not necessarily related to the squeezing parameter). Then, 
\begin{equation}
    \rho _S^{12}(t)=\sum_{n=0}^\infty  P_n  \Tr_{E}(U(t) \rho_S \otimes  \Pi_n^1\otimes  \Pi_n^2 U^\dagger(t) )
\end{equation}
with $  P_n= (1-u^2) u^{2n}$. Now supposing the initial system state is 
\begin{equation}\label{262626}
    \ket{\psi_S}=\sum_{i,j=0}^1a_{ij}\ket{ij}  
\end{equation}
 we consider the time-evolution of $\rho_S^{12}(t)$,
    \begin{equation}\label{E1}
         \rho _{S}^{12}(t)=\left(
            \begin{array}{cccc}
                \abs{a_{11}}^2  & a_{11} a_{10}^*  \tilde{k}_2(t)  & a_{11} a_{01}^* \tilde{k}_1(t) & a_{11} a_{00}^*  k_{12}(t)\\
              & \abs{a_{10}}^2  & a_{10} a_{01}^*  \Lambda_{12}(t) & a_{10} a_{00}^* k_1(t)  \\
               &   &\abs{a_{01}}^2 & a_{01} a_{00}^* k_2(t)   \\
             c.c.  &   &   & \abs{a_{00}}^2 \\
            \end{array}
            \right) ,
    \end{equation}
  where
    \begin{align}\label{allphase}
        k_1(t)= &e^{-2i \epsilon_1 t}\prod_k(\sum_{n=0}^\infty  P_n\chi^{1000}_{k,n}), \notag\\
         k_2(t)= &e^{-2i \epsilon_2 t}\prod_k(\sum_{n=0}^\infty  P_n\chi^{0100}_{k,n}),\notag \\
        \tilde{k}_1(t)=& e^{-2i \epsilon_1 t}\prod_k(\sum_{n=0}^\infty  P_n\chi^{1101}_{k,n}), \notag\\
         \tilde{k}_2(t)= &e^{-2i \epsilon_2 t}\prod_k(\sum_{n=0}^\infty  P_n\chi^{1110}_{k,n}),\notag \\
        k_{12}(t)= &e^{-2i (\epsilon_1+\epsilon_1) t}\prod_k(\sum_{n=0}^\infty  P_n\chi^{1101}_{k,n}), \notag\\
         \Lambda_{12}(t)= &e^{-2i (\epsilon_1-\epsilon_1) t}\prod_k(\sum_{n=0}^\infty  P_n\chi^{1110}_{k,n}),\notag \\
    \end{align}
   and
    \begin{equation}\label{chio}
        \chi^{nmrs}_{k,i}=\Tr\Bigl(\Pi_i^1\otimes  \Pi_i^2 \exp[\sum_{j=1}^2\gamma_{k,nmrs}^j(t) {b_k^j}^\dagger-H.c.]\Bigr),
    \end{equation}
    with
    \begin{align*}
        \gamma_{k,nmrs}^1(t)=&((-1)^n-(-1)^r)\beta_k^1(t),\notag \\
        \gamma_{k,nmrs}^2(t)=&((-1)^m-(-1)^s)\beta_k^2(t).
    \end{align*}
    The analytic evaluations of these phase factors in \cref{allphase} can be fund in \cref{AnnE}.
    
\begin{figure}[t]
    \centering
    \includegraphics[width=0.45\textwidth]{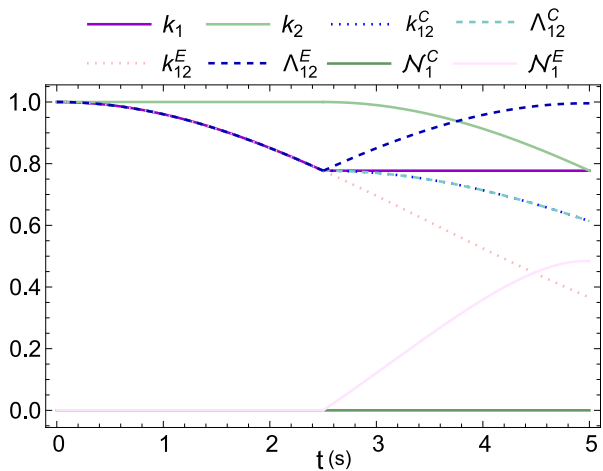}
    \caption{The time-evolutions of the phase factors in \cref{allphase} and of the non-Markovianity. They are dimensionless numbers. Two cases are considered: entangled sub-environments (E) and the classically correlated sub-environments (C). Here we take the parameters $\alpha_{1,2}=1$, $\omega_c=10^{-2}$, $r=3$, $t_1^s=0$, $t_1^f=t_2^s=2.5$, and $t_2^f=5$.}\label{NMFig}
    \end{figure}

     Numerically, we take the continuum limit as in \cite{nlm} for the spectral density $J_j=\alpha_j \omega \exp(-\omega/\omega_c)$ with equal cutoff frequency $\omega_c$ but different couplings $\alpha_j$ for the two bosonic baths.
    We  then obtain in \cref{NMFig} the time evolutions of these phase factors, both for the case with classical correlations and for the entangled case. It shows that the classically correlated environment cannot give rise to 
    the nonlocal memory effect, while the entangled environment can. And the phase factor $ \Lambda_{12}$ is the key difference.
    
We find that the time-evolutions of $k_1$ and $k_2$ in two cases are the same, so we didn't distinguish them in  \cref{NMFig}.  And only the evolutions of $  k_{12}$ and $  \Lambda_{12}$ are different. Moreover,
    when $t\leq 2.5$, the time-evolutions of $  k_{12}$ and $  \Lambda_{12}$ are also the same in these two case; but after $t=2.5$, the two cases differ. In particular, $  \Lambda_{12}$ is decreasing in the case with classical correlation. In other words, in the case with entangled sub-environments,  some phase factor can be recovered (as they will increase after $t=2.5$), so that the non-Markovianity $\mathcal{N}_1^E$ becomes non-trivial. While in the case with classically-correlated sub-environments, none of the phase factor can be recovered, so $\mathcal{N}_1$ keeps zero.
  Notice that in      \cref{NMFig}, we have taken the optimal state as
\begin{equation}\label{ops}
    \ket{\psi}_{AS_1S_2}=\frac{1}{\sqrt{2}}(\ket{01}_S\otimes\ket{0}_A+\ket{10}_S\otimes\ket{1}_A),
\end{equation}
so that the coefficients $a_{ij}$ of the initial system state have been determined.

\begin{figure}[t]
      \centering
      \includegraphics[width=0.45\textwidth]{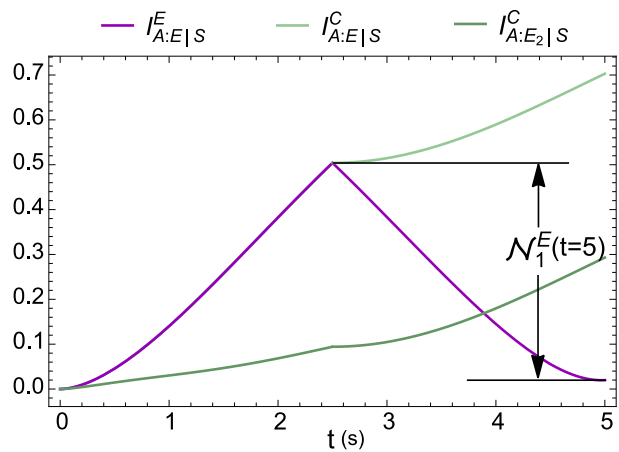}
      \caption{The time-evolutions of the conditional mutual information in two cases (E and C). They are dimensionless numbers. The parameters are the same as in \cref{NMFig}.}\label{CMIFig} 
      \end{figure}

In \cref{CMIFig}, we show the time-evolutions of the leaked information in these two cases in the same setup. For $t\leq 2.5$, the total leaked information in two cases coincide. After $t=2.5$, we see the information backflow in the case of entangled sub-environments, but not in case of classically-correlated sub-environments. The increase of $I^C_{A:E_2|S}$ imply that  the classically correlation in the environment may not give rise to the nonlocal memory effect.
Moreover, from the evolution of $I^C_{A:E_2|S}$, we see that  the sub-environment $E_2$ can get leaked information as long as  there are correlations between sub-environments, no matter what type of correlation is.  But,  in the entangled case it is easier to generate the  leaked information in $E_2$. For instance,  from \cref{CMIFig} we see that $I^C_{A:E_2|S}(t=2.5)$ is much smaller than $\mathcal{N}_1^E$; but because  $\delta I(A:E_1E_2|S_1S_2)=\delta I(A:E_2|S_1S_2)$ under the  unitary $U_{S_2E_2}$, we have
\begin{equation}
    \mathcal{N}_1^E=I^E_{A:E_2|S}(t=2.5)-I^E_{A:E_2|S}(t=5),
\end{equation}
which proves that $I^C_{A:E_2|S}(t=2.5)$ is much smaller than $I^E_{A:E_2|S}(t=2.5)$. This means the leaked information in sub-environment $E_2$  are totally different in two cases for $t\leq 2.5$. And the decrease of  $I^E_{A:E_2|S}$ during $t \geq 2.5$ show that  there is nonlocal memory effect in the entangled case.

Be careful that the nonlocal memory effect here is different from the non-local leaked information in \cref{s2a}. The leaked information $I_{A:E_2|S}$ is nonvanishing for both cases at $t>0$. Only the entangled case gives rise to the nonlocal memory effect. But,  both cases do not give rise to any non-local leaked information. The reason is as follows. On the one hand, evolution $U_{S_1E_1}$ can not bring any non-local correlation of $A:E_2$ for state $\rho_{AS_1S_2}\otimes\rho_{E_1E_2}$. Hence, the leaked information $ I(A:E_2|S)$ is non-resourceful. On the other hand, since $\delta I(A:E_1E_2|S_1S_2)=\delta I(A:E_1|S_1S_2)$ under the  unitary $U_{S_1E_1}$, the leaked information $ I(A:E_2|S)$ should be redundant according to \cref{19}.

 \section{Conclusion and outlook}\label{CO}
In this paper, we have discussed an equivalent form of the LFS non-Markovianity measure by using  quantum conditional mutual information. 
 We first find that the LFS  measure, using the telescopic relative entropy as distinguishability measure, doesn't detect more non-Markovian cases than a  BLP measure. 
 Then we show that the new form of the LFS measure $\mathcal{N}_1$ in terms of quantum conditional mutual information can give rise to the definition of leaked information for structured environment.
The leaked information defined here lifts the  quantum conditional mutual information as a bound on the deviation from Markovianity \cite{FR15,SW15} to a quantity directly related to the (LFS) measure of non-Markovianity.  The leaked information is exploited here to show that the environment with classically correlated sub-environments
 still cannot generate nonlocal memory effect, which  suggests that there may be some deeper relationships  between entangled environment and nonlocal memory effect.

It is interesting  that the classical correlation contained in the leaked information share some common features with the structured environment as studied in Quantum Darwinism, e.g. the redundancy of classical information \cite{BZ06}. Using the leaked information, we look forward to studying quantitatively the relation between the saturation of Quantum Darwinism and the difficulty of the backflow of leaked information, which is a general result reached by various recent works (see, e.g. \cite{LPP19} and references therein).

In many circumstances the non-Markovianity will be small if the environment becomes very large. For example, in \cite{FMP19}, it is shown by using random unitaries that almost all open quantum processes will concentrate on the Markov case, when the environment is large enough. This almost Markovian phenomenon can be intuitively understood from the perspective of local propagation of information (cf. \cref{LEII}), or from the bounds on almost Markov chains \cite{FR15,SW15}. The leaked information introduced above allows us to quantitatively study this phenomenon. 

Finally,  we remark that the non-Markovianity measure such as the RHP, BLP and LFS measures have the common problem that they are sufficient but not necessary conditions for characterizing non-Markovianity. However, in the formalism of process tensor one has a necessary and sufficient condition for Markovian quantum process  \cite{PRFPM18}. The process tensor formalism can also describe multi-time observables, so that some examples showing the unnecessity of the above non-Markovianity measures can be unambiguously characterized \cite{TPM19}. It is interesting to investigate the quantification of quantum non-Markovianity in such multi-time or process framework also using the (multipartite) quantum conditional mutual information. 
We hope to return to these topics in future investigations.
\begin{acknowledgments}
    We thank the referees for their helpful comments and suggestions that significantly polish this work. We also thank G. Karpat for helpful comments. ZH  is supported by the National Natural Science Foundation of China under Grant Nos. 12047556, 11725524 and 61471356.
\end{acknowledgments}

\appendix
\section{Derivation of \eqref{99}}\label{AnnB}
We present some details about \eqref{99}:
\begin{widetext}
\begin{align*}
       S(\rho_{SA}||\rho_S\otimes\rho_A)=&\frac{1}{2}\Tr_{SA}\Bigl[(\rho^1_S\otimes \Pi^1_A+\rho^2_S\otimes \Pi^2_A)\Bigl(\log\frac{1}{2}(\rho^1_S\otimes \Pi^1_A+\rho^2_S\otimes \Pi^2_A)-\log \frac{1}{4}(\rho^1_S+\rho^2_S)\otimes (\Pi^1_A+\Pi^2_A)\Bigr )\Bigr]
       \notag\\
       =&  \frac{1}{2}\Tr_{SA}\Bigl[\rho^1_S\otimes \Pi^1_A\Bigl(\log \rho^1_S\otimes \Pi^1_A-\log \frac{1}{2}(\rho^1_S+\rho^2_S)\otimes \Pi^1_A\Bigr)\Bigr]+ \notag\\
       &+\frac{1}{2}\Tr_{SA}\Bigl[\rho^2_S\otimes \Pi^2_A\Bigl(\log (\rho^2_S\otimes \Pi^2_A)-\log \frac{1}{2}(\rho^1_S+\rho^2_S)\otimes \Pi^2_A\Bigr)\Bigr]
       \notag  \\
       =&\frac{1}{2}\Bigl[S(\rho^1_S||\frac{\rho^1_S+\rho^2_S}{2})+S(\rho^2_S||\frac{\rho^1_S+\rho^2_S}{2})\Bigr]
       =\frac{-\log (1/2)}{2}\Bigl[S_{\frac{1}{2}}(\rho^1_S||\rho^2_S)+S_{\frac{1}{2}}(\rho^2_S||\rho^1_S)\Bigr]
\end{align*}
\end{widetext}
where the first equality is the definition of quantum relative entropy. In the second equality, we have used the linearity of trace and the orthogonality $\Pi_A^1\Pi_A^2=0$ after
 expanding the logarithmic functions $\log(A+B)$ into series. In the third equality, we have discarded the $A$-part of the tensor product in the relative entropies.
 
 \section{A new  non-Markovianity measure }\label{BBB}
According to the relation between the RHP measure and the BLP measure, we can generalize $  \mathcal{N}_1$ to a new measure $  \mathcal{N}_2$ \eqref{NMMBHP} which is related to both the RHP measure and the BLP measure. 

The RHP measure for quantum non-Markovianity can be realized in the way of the BLP measure, if we add a suitable ancillary  $A'$ to the open system $S$ in such a way that the CP-divisibility condition can be  recovered \cite{BJA17}.
The corresponding  non-Markovianity measure can be written as
\begin{align}
    \mathcal{N}_{\text{RHP}}(\Lambda)=\sup_{\rho,\tau}&\int_{\frac{d}{dt}D(\mathbbm{1}_{A'}\otimes\Lambda_t\rho,\mathbbm{1}_{A'}\otimes\Lambda_t\tau)>0}\notag \\
    &\frac{d}{dt}D(\mathbbm{1}_{A'}\otimes\Lambda_t\rho,\mathbbm{1}_{A'}\otimes\Lambda_t\tau)dt,\label{RHP}
\end{align}
where $\rho, \tau\in \mathcal{B}(\mathcal{H}_S\otimes\mathcal{H}_{A'})$.
The primed ancillary $A'$ could be understood as an copy of the system $S$, if the extended dynamical map $\mathbbm{1}_{A'}\otimes\Lambda$ in defining the CP condition comes from the Choi-Jamio\l kowski isomorphism. But in  \cite{BJA17} it is proved that the CP-divisibility can be formulated as a distinguishability condition, if $A'$ is extended to be of $\dim S+1$ dimensions.

Here we still work in the ``system+ancillary+environment'' setup, but consider the system $S$ to be extended to $SA'$ with $\dim A'=\dim S+1$, as constructed in \cite{BJA17}.
Given this, we propose the following new non-Markovianity measure as an extension of the measure $\mathcal{N}_1(\Lambda)$ \eqref{NMM} and also $\mathcal{N}_{\text{RHP}}$ \eqref{RHP},
\begin{equation}\label{NMMBHP}
    \mathcal{N}_2(\Lambda):=  \sup_{\rho_{SA'AE}}\int_{\frac{d}{dt} I(A:E|SA')<0} \abs{ \frac{d}{dt} I(A:E|SA')}dt.
\end{equation}
Comparing this $\mathcal{N}_2$ to $\mathcal{N}_1$, we see the replacement $\rho_S \to \rho_{SA'}$, and $\mathcal{N}_2$ reduces to $\mathcal{N}_1$ if $A'$ is  trivial. Since $\mathcal{N}_2$ is an extension of $\mathcal{N}_1$, $\mathcal{N}_2$ can in principle detect more non-Markovianity than $\mathcal{N}_1$. It is easy to see that $\mathcal{N}_2(\Lambda)$ detects more non-Markovian cases  than $\mathcal{N}_{\text{RHP}}(\Lambda)$.

\section{Quantum conditional mutual information, recovery map and Markovianity}\label{AnnA}
The quantum conditional mutual information plays an important role in state reconstructions. For a tripartite quantum system $A\otimes B\otimes C$, the total state $\rho_{ABC}$ can be reconstructed from the bipartite reduction $\rho_{AB}$ through a quantum operation $B\rightarrow B\otimes C$, if the quantum conditional mutual information $I(A:C|B)=0$ \cite{HJPW04}. When $I(A:C|B)\neq0$, the total state still can be approximately reconstructed by a recovery  channel $\mathcal{R}_{B\to BC}\equiv\mathcal{R}$ such that $\mathcal{R}\rho _{AB}=\sigma _{ABC}$. The difference, e.g. trace distance $D(\sigma,\rho)$,  between
$\sigma _{ABC}$ and the proposed $\rho_{ABC}$ is bounded by the conditional mutual information
$I(A:C|B)$ \cite{FR15},
\begin{equation}\label{AS2ASEUI}
   D(\sigma_{ABC},\rho _{ABC})^2 \leqslant \ln2 I(A:C|B)
\end{equation}
This bound \eqref{AS2ASEUI} corroborates the above-mentioned result that if $I(A:C|B)=0$, then one can recover exactly the  total state $\rho_{ABC}$.

Conversely, if we can reconstruct the $\rho_{ABC}$ from  $\rho_{AB}$, then $I(A:C|B)=0$. Indeed, the quantum conditional mutual information can be rewritten in terms of the conditional entropies as
\begin{equation}
    I(A:C|B)=S(\rho_{A}|\rho_{B})-S(\rho_{A}|\rho_{BC}).
\end{equation}
Then by the data processing inequality, one has
\begin{equation}
    I(A:C|B)\leqslant S(\sigma_{A}|\sigma_{BC})-S(\rho_{A}|\rho_{BC}),
\end{equation}
the right-hand sight of which can be bounded by the trace distance $D$ \cite{FR15,AF04},
\begin{equation}\label{RACMI}
    I(A:C|B)\leqslant 7\log_2(\dim A)\sqrt{D(\rho_{ABC},\sigma_{ABC})}.
\end{equation}
When $D(\rho_{ASE},\sigma_{ABC})=0$, one has $I(A:C|B)=0$. A special case is when there is no system-environment correlation, e.g. $\varrho_{SE}=\varrho_{S}\otimes\rho_E^0$, one has $D(\rho_{ASE},\mathcal{R}_{S\to SE}\rho _{AS})=I(A:E|S)=0$.

In the ``system-ancillary-environment'' setup, if initially $I(A:E|S)=0$, then the dynamical change of $I(A:E|S)$ must have the following property
\begin{equation}\label{C5}
    \lim_{t\to 0}\frac{d}{dt} I(A:E|S)\geqslant0,
\end{equation}
since $I(A:E|S)\geqslant0$. In other words, the initial dynamical evolution must be Markovian.

Suppose the system is interacting with two environments $E_1$ and $E_2$. If initially
$I(E_2:A|SE_1)=0$, then one has the initial evolution  $\rho_{SE_1}(t)=\Lambda_{t,t_0}\rho_{SE_1}(t_0)$ with $\Lambda_{t,t_0}$ being a CPTP map. If furthermore $I(E_1:A|S)=0$ initially, then $\rho_{S}(t)=\Tr_{E_1E_2}\Lambda_{t,t_0}\mathcal{R}_{S\to SE_1}\rho_{S}(t_0)$, where $\Lambda'\equiv\Tr_{E_1E_2}\Lambda_{t,t_0}\mathcal{R}^{P}_{S\to SE_1} $ is still a CPTP map. This is consistent with the chain rule of the conditional mutual information
\begin{equation}\label{B6}
    I(E_1E_2:A|S)=I(E_1:A|S)+I(E_2:A|SE_1).
\end{equation}

We also need the notion of {\it recoverability} which for the purpose of this paper is roughly the fidelity between the original state $\rho_{ABC}$ and the recovered state $\sigma_{ABC}$ obtained by the recovery map. More precisely, the fidelity of recovery is defined as the optimized fidelity of the recovery channel, $F(A:B|C)_\rho=\sup_\mathcal{R}F(\rho_{ABC},\sigma_{ABC})$ \cite{SW15}, where $F(\rho,\sigma)=\norm{\sqrt{\rho}\sqrt{\sigma}}_1^2$ id the fidelity between two quantum states.

 \section{Local expansion and leaked information in sub-environment}\label{LEII}
We have pointed out in the main text that the quantum conditional mutual information $I(A:E|S)$ can quantify the amount of the leaked information. Here we study the leaks from the point of view of localized propagation of information (i.e. the Lieb-Robinson bounds).

Suppose the ``S+A+E'' setup is defined on a lattice, then the influence of $\rho_{SA}$ on E is localized and bounded by the  Lieb-Robinson bound in the entropic form \cite{IKS17}
 \begin{equation}\label{LR}
     \abs{S(\rho_{SA}(t))-S(\rho^R_{SA}(t))}\leqslant C e^{-\alpha(d-v_\text{LR}(t-t_0))}
 \end{equation}
where $C,\alpha$ are a constant, $d$ is the lattice distance and $v_{\text{LR}}$ is the Lieb-Robinson velocity.   Here $\rho_{SA}(t)=\Tr_E(e^{-iH(t-t_0)}\rho_{SEA}(t_0)e^{iH(t-t_0)})$ with $H=H_{SA}+H_{{SA}{E}_d}+H_{{E}_d}+H_{{E}_d \overline{{E}_d}}+H_{\overline{{E}_d}}$; $E_d$ denotes the part of environment that directly interacts with the system. On the other hand, $\rho^R_{SA}(t)=\Tr_E(e^{-iH^R(t-t_0)}\rho_{SEA}(t_0)e^{iH^R(t-t_0)})$ with $H^R=H_{SA}+H_{{SA}{E}_d}+H_{{E}_d}$. $H^R$ is the Hamiltonian with the noncontributing part $\overline{{E}_d}$ of the environment discarded; this discarded part could affect the system only after the time $t\sim d/v_\text{LR}$. By \eqref{LR}, we have an inequality for the mutual information
 \begin{equation}
     \abs{I_{A:S}(\rho_{SA}(t))-I_{A:S}(\rho^R_{SA}(t))}\leq  2 C e^{-\alpha(d-v_{LR}(t-t_0))}.
 \end{equation}
Since $U^{R}(t)=e^{-i(H_\mathcal{S}+H_{\mathcal{S}\mathcal{E}_d}+H_{\mathcal{E}_d})(t-t_0)}$ does not change $I(A:E_dS)$, we obtain
 \begin{equation}
     I_{A:S}(\rho^R_{SA}(t))=  I_{A:E_d|S}(\rho^R_{SEA}(t))+I_{A:E_dS}(\rho_{SEA}(t_0)).
 \end{equation}
All in all, we have
 \begin{align}\label{FIBF}
     \abs{I_{A:S}(\rho_{SA}(t))-I_{A:S}(\rho_{SA}(t_0))}\leqslant  2 Ce^{-\alpha(d-v_{LR}(t-t_0))} \notag\\
  +\abs{  I_{A:E_d|S}(\rho^R_{SEA}(t))-  I_{A:E_d|S}(\rho^R_{SEA}(t_0))},
 \end{align}
 which shows that the quantum conditional mutual information can be used to quantify the deficit part of local propagation of information.

 \section{The phase factors}\label{AnnE}
Letting
    \begin{equation}\label{optrans}
        \hat{q}_j=\frac{1}{\sqrt{2}}(\hat{b}_j+\hat{b}_j^\dagger), \quad \hat{p}_j=\frac{-i}{\sqrt{2}}(\hat{b}_j-\hat{b}_j^\dagger),
    \end{equation}
    we have 
    \begin{widetext}
    \begin{align}
        \Tr\Bigl(\Pi_n \exp[-i x\hat{p}]\Bigr)=\exp\Bigl(-\frac{x^2}{4}\Bigr)\Bigl(1- \frac{n}{2}x^2
        + \frac{n(n-1)}{16}x^4-\dots+(-1)^i\frac{C^n_i}{2^{i}i!} x^{2i} +\dots \Bigr)
        =\exp\Bigl(-\frac{x^2}{4}\Bigr)L_n\Bigl(\frac{x^2}{2}\Bigr).\label{PMOFH}
    \end{align}
   where $L_n$ is the  Laguerre polynomial. Using this \cref{PMOFH} and the Hardy-Hille formula, we obtain
   \begin{align}
   \Tr\Bigl( \rho_u\exp[-i x_1\hat{p}_1-i  x_2\hat{p}_2] \Bigr)
   =\exp\Bigl(-\frac{\cosh(2r)}{4}(x_1^2+x_2^2)\Bigr)\times I_0\Bigl(\frac{(x_1^2x_2^2u^2)^{1/2}}{1-u^2}\Bigr), \label{LPHH}
\end{align}
where $ I_\alpha(x)$ is the Modified Bessel functions of first kind. The similar formulas can be obtained for the expectation of the momentum shift operator. Then combining \cref{modin,chio,optrans,PMOFH,LPHH}, we have
\begin{align}\label{finalcal}
    \prod_k( \sum_{i=0}^\infty  P_i \chi^{nmrs}_{k,i})=\exp\Bigl\{\int_0^\infty d k \Bigl(-\frac{\cosh(2r)}{4}f_k^{nmrs}
    + \log\bigl[ I_0(\frac{\abs{g_k^{nmrs}} \sinh(2r)}{2})\bigr]\Bigr) \Bigr\},
\end{align}
with
\begin{align*}
    f_k^{nmrs}=2(\abs{\gamma_{k,nmrs}^1(t)}^2+\abs{\gamma_{k,nmrs}^2(t)}^2), \quad
    g_k^{nmrs}=2\abs{\gamma_{k,nmrs}^1(t)\gamma_{k,nmrs}^2(t)}.
\end{align*}
From \cref{finalcal}, we can evaluate \cref{E1} and its time-evolution. 
\end{widetext}

\end{CJK*}

\end{document}